\documentclass[11pt]{article}
\usepackage{graphicx}
\usepackage{dcolumn}
\usepackage{bm}
\usepackage{float}
\usepackage{placeins}
\usepackage{hyperref}
\usepackage{natbib}

\usepackage{subfiles}

\usepackage[T1]{fontenc}
\usepackage[utf8]{inputenc}
\usepackage{authblk}

\usepackage{amsmath}
\usepackage{relsize,exscale}
\DeclareUnicodeCharacter{2212}{\textendash}

\title{A Study of Cosmological Models Based on Long Gamma-ray Bursts Duration Histogram}

\author[1]{Zeinab Kalantari\thanks{z.kalantari@sharif.edu}}
\author[1]{Sohrab Rahvar}
\affil[1]{Department of Physics, Sharif University of Technology, Tehran 11365-9161, Iran}

\begin{document}
\maketitle
\begin{abstract}
The duration of more than one thousand gamma-ray bursts (GRBs) has been measured by Swift satellite. Besides the redshift distribution of GRBs, the burst duration could be another significant property of GRBs that can be analyzed. In this project, First, we find the detection rate of Swift/BAT for a cosmological model with the $ \omega CDM$ model, then by performing a Monte-Carlo simulation, we find the "long" GRB duration histogram to compare the cosmological model with the $ \omega CDM$ model and the $\Lambda CDM$ model. The $\chi^2$ minimization method is employed to determine the optimal value of $\alpha$ in the dark energy equation of state, $P= c \omega(z) \rho$, with $\omega(z)=1+\alpha z$ as $\alpha=0.08 ^{+ 0.04}_{-0.02}$. We showed that the $\omega CDM$ model is statistically favored over the $\Lambda CDM$ model.   Other studies by measuring the expansion rate of the Universe directly from the SNIa data by Pantheon and DES-SN3YR samples and the combinations of SNIa, CMB, and BAO  data confirmed that the value of $\alpha$ in the dark energy equation of state is not zero \citep{Dark_energy_Abbott, 2011Beutler, 2014Anderson, 2015Ross, 2016PlanckCollaboration, Alam2017, Mazumdar2021}.




\begin{description}
\item[Keywords]
Gamma-Ray Bursts, Dark Energy, Cosmological Models
\end{description}
\end{abstract}
\maketitle

\section{Introduction}
\label{Sec:introduction}
Gamma-ray bursts represent the most luminous electromagnetic bursts in the universe which radiate isotropic equivalent energy between  $10^{48}-10^{55}$ erg \citep{Kulkarni}.
Based on the gamma-ray burst duration time scale, GRBs are typically categorized into two groups: long gamma-ray bursts, which have a duration of more than 2 seconds ($T_{90}>2$), and short gamma-ray bursts, with a duration less than 2 seconds ($T_{90}<2$) \citep{Kouveliotou}.
The prevailing hypothesis is that long gamma-ray bursts (LGRBs) occur when a massive star experiences core collapse \citep{Fryer}. In contrast, the sources of short gamma-ray bursts are believed to result from mergers of two neutron stars or merging a neutron star and a black hole in a compact binary system \citep{Berger}. 
 
Gamma-ray bursts' powerful luminosity enables us to explore the universe at the farthest redshift reaches ever explored. The maximum observable redshift of gamma-ray bursts is estimated to be around $z=9.4$ \citep{Cucchiara}, so GRBs could serve as an excellent tool to investigate the characteristics of the cosmos at high redshifts, such as constraining the cosmological parameters such as dark energy equation of state.
\par
From the late 1990s, observations of high-redshift Type Ia supernovae (SNIa) unveiled a crucial fact that the Universe is currently undergoing accelerated expansion \citep{Perlmutter}. The common assumption is that the observed accelerated expansion is attributed to dark energy with negative pressure and is related to energy density by the dark energy equation of state ($P= c \omega(z) \rho$). In the case of $\Lambda CDM$ cosmology $\omega=-1$, other alternative models of dark energy proposed that $\omega$ is not constant and depends on the redshift $z$. In this work, we take the simplest extension of $\omega CDM$ model where $\omega =-1 +\alpha z$ and compare the duration histogram of LGRBs resulting from this $\omega CDM$ with standard model of cosmology.

\par
Earlier studies focused on using Gamma-Ray Bursts (GRB) for determining cosmological parameters (such as the fundamental characteristics and evolution of dark energy) relied on methods involving the calibration of the peak photon energy of GRBs. Like Type Ia supernovae, using GRB correlations has been suggested as a way to standardize their energies and/or luminosities. Numerous GRB correlations have been put forward for cosmological purposes.
\citep{Wang, Demianski, Hanbei, Fenimore, Norris, Amati, Ghirlanda, Yonetoku, Liang, Firmani, Ghirlanda2, Dainotti, Tsutsui, Izzo, Hu, Wang, Dainotti, Dainotti2, Dainotti3, Li, Dai}. By employing the GRB luminosity correlations as a cosmological standard candle, similar to the approach used with Type Ia supernovae (SN), Gamma-ray bursts (GRBs) have the potential to extend the Hubble diagram to higher redshifts \citep{Schaefer}. However, finding a direct relationship between luminosity and distances for GRBs is challenging due to a limited understanding of the mechanism of GRBs progenitors. In other words, GRB correlations lack a clear underlying physics, in contrast to the well-defined Chandrasekhar limit in the physics of Type Ia supernovae (SNe Ia). Therefore, the GRB candle is considerably less standardized than the SN Ia candle. On the other hand, achieving a robust GRB calibration, as seen in the case of Type Ia supernovae (SNe Ia), needs a low-redshift sample of GRBs. However, nearby GRBs typically exhibit much lower luminosity (see $E_{iso}$ and redshift distribution for GRBs in Fig.1 of \citep{Lan2023})
 Consequently, GRBs may not be used as standard candles \citep{Lin, Khadka}.
 \\
In this study, we used data from the BAT telescope on the board of the Swift satellite to obtain the duration of long GRBs. The Burst Alert Telescope (BAT) is a highly sensitive, wide-field telescope. This instrument is specifically designed to monitor a significant portion of the sky for the detection of GRBs. The BAT provides both the burst trigger and the accurate position of about 4 arcminutes \citep{Barthelmy2005}. Then two other instruments of Swift, the XRT (X-ray telescope) and the UVOT (ultraviolet and optical telescope) follow up on the BAT burst positions to measure the redshift of GRB by monitoring counterparts of GRBs in other wavelengths.

\par
The paper is organized as follows. Section \ref{Sec:Theoretical} presents the theoretical calculation of LGRBs rate. Section \ref{Sec:Efficiency} describes how to find Swift/BAT detection rate for the $\Lambda CDM$ model and the cosmological model with the dark energy model of $\omega(z)=-1+\alpha z$. In section \ref{Sec:GRB_Histogram} with an example, we present our method for numerical calculation of the observational rate of long GRBs to find LGRBs duration histogram. In section \ref{Sec:w} we use a $\chi^2$ minimization method to explore the best $\alpha$ for the dark energy model.



\section{Theoretical calculation of long GRBs rate}
\label{Sec:Theoretical}

\begin{figure}[!htbp]
\centering
\includegraphics[width=0.7\linewidth]{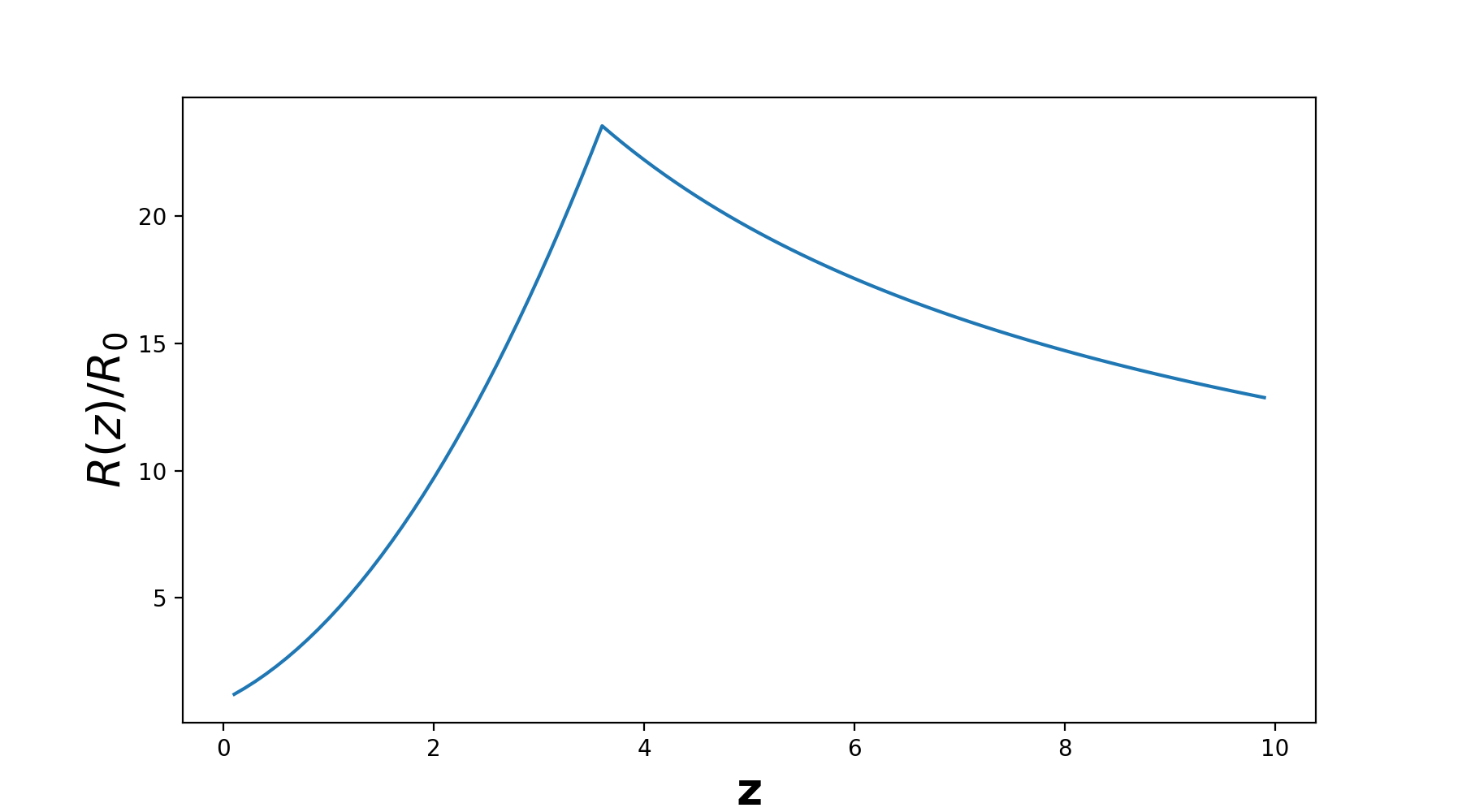}
\caption{\label{Fig_R_Z}
The comoving rate of LGRBs versus redshift ($R(z)$) per local rate of LGRBs ($R_0$) based on Eq.(\ref{Eq:R_Wanderman})
}
\end{figure}

The cosmic rate of LGRBs in the comoving frame is defined as
\begin{equation}
    R(z)= \frac{dN}{dV \ dt_{Source}}
    \label{Eq:R_comoving}
\end{equation}
in this equation, $N$ is the number of LGRBs we observe at the cosmological redshift of $z$ and $V$ is the comoving volume, and $dt_{Source}$ is the time interval at the source (GRBs) frame. 
Long-duration gamma-ray bursts (LGRBs) are believed to be the result of the death of massive stars, which may have low metallicity \citep{Woosley2006} and have been linked to Type Ic supernovae through observations \citep{Hjorth2003}. So the rate of LGRBs could related to the star formation rate.
Similar to most studies, here we use \cite{Wanderman_2010} comoving rate function of Swift's gamma-ray bursts based on the assumption that GRBs rate  generally follows the shape of the cosmic star-formation rate for redshifts $z\approx 3$, the rate decreases less steeply at higher redshifts compared to the rapid decline in the star-formation rate beyond redshift of $z>4$
\begin{equation}
 R(z)= R_0
\begin{cases} 
      (1+z)^{\nu_1} & z < z_{*} \\
      (1+z_*)^{(\nu_1-\nu_2)}(1+z)^{\nu_2} & z_{*} < z
\label{Eq:R_Wanderman}
\end{cases}
\end{equation}
where $R_0 = 0.84 \ (Gpc^{-3} yr^{-1})$ is local rate of GRBs, $z_* = 3.6$, $\nu_1 = 2.07 $ and $\nu_2 = -0.70$ match best
with the characteristic of observed GRBs \citep{Lien2014}. The schematic of Eq. (\ref{Eq:R_Wanderman}) is shown in Fig.\ref{Fig_R_Z}.
\par
\begin{figure}[!htbp]
\centering
\includegraphics[width=0.7\linewidth]{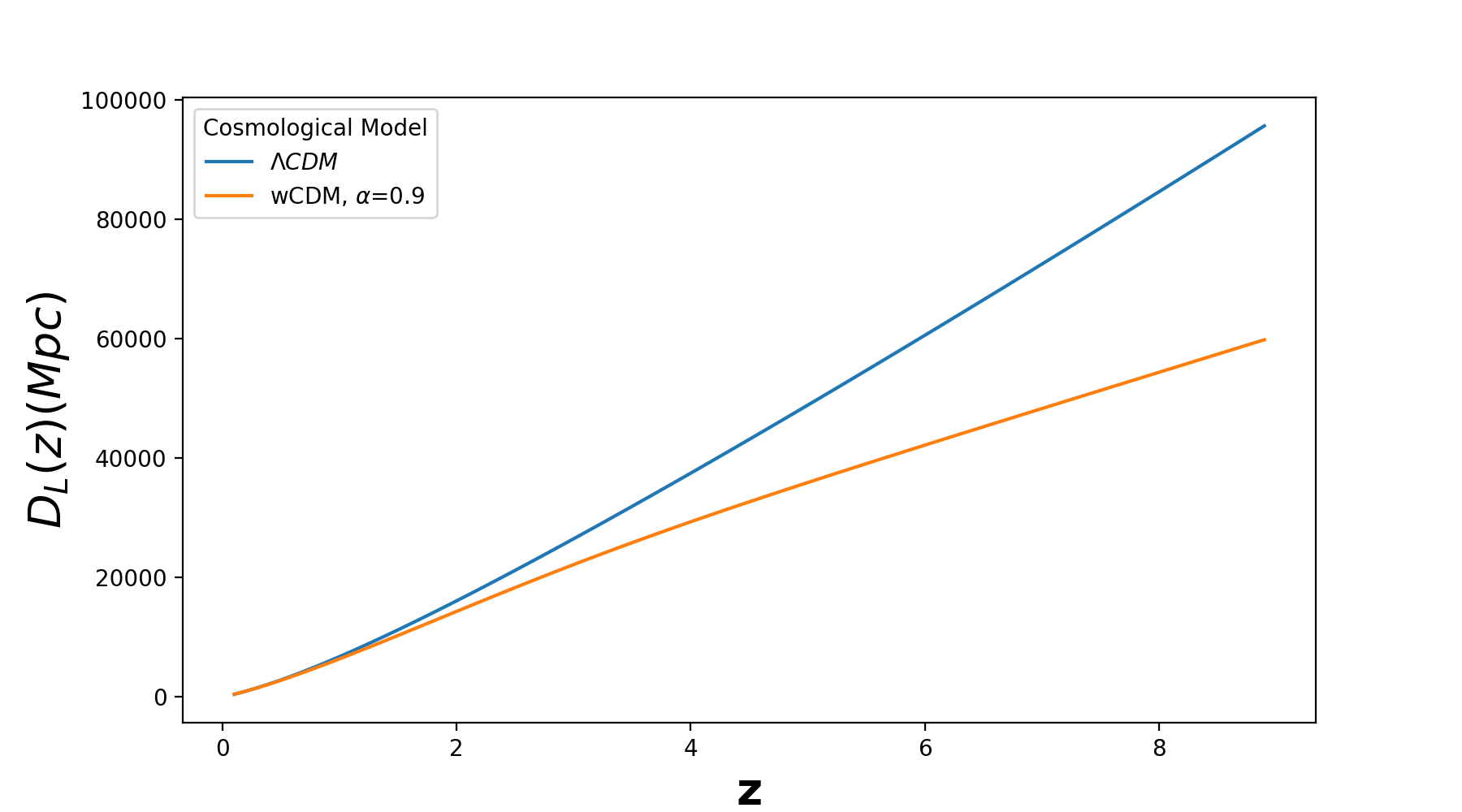}
\caption{\label{Fig_D_C} 
Difference between $ D_{C} (z)$ of $\omega CDM$ model with $\alpha=0.9$ (orange) and $\Lambda CDM$ model (blue)
(see Eq.\ref{Eq:ComovingDistFinal})
}
\end{figure}

The observed GRBs rate on Earth is usually defined as
\begin{equation}
    \Gamma(z)= \frac{dN}{d \Omega \ dz \ dt_{Obs}}
    \label{Eq:Gamma1}
\end{equation}
substituting Eq.(\ref{Eq:R_comoving}) in above equation gives observed GRBs rate $\Gamma$ as
\begin{equation}
    \Gamma(z)= \frac{dN}{ d \Omega \ dz \ dt_{Obs}} =  R(z) \frac{dV}{ d \Omega \ dz} \frac{dt_{Source}}{dt_{Obs}} =  \frac{R(z)}{1+z}  \frac{dV}{ d \Omega \ dz}
    \label{Eq:Gamma2}
\end{equation}
the $1/(1+z)=dt_{Source}/dt_{Obs}$ factor in Eq.(\ref{Eq:Gamma2}) equation is due to cosmological time dilation. Comoving volume $dV$ equals to
\begin{equation}
    dV(z)= d \Omega D_{C}(z)^2 dD_{C}(z)
    \label{Eq:dV_comov}
\end{equation}
were the comoving distance at the cosmological redshift of $z$ is defined as
\begin{equation}
    D_{C} (z)=\int_0 ^z \frac{c\ dz'}{H(z')}
    \label{Eq:ComovingDist}
\end{equation}
therefore, from Eq.(\ref{Eq:dV_comov}), Eq. (\ref{Eq:ComovingDist}) and Eq.(\ref{Eq:Gamma2}) the $\Gamma$ parameter would be
\begin{equation}
    \Gamma (z)=\frac{R(z)}{(1+z)}\frac{D_{C}(z)^2 dD_{C}(z)}{ dz} = \frac{c\ R(z)}{(1+z)} \frac{D_{C}(z)^2} {H(z)}
    \label{Eq:Gamma_Def}
\end{equation}
From Friedmann's equations, we can find the Hubble parameter for cosmological models. 

\begin{equation}
\begin{cases} 
        H^2=(\frac{\dot{a}}{a})^2=\frac{8 \pi G}{3} \rho+\frac{c^2\Lambda}{3}= H_0^2(\Omega_M(z)+\Omega_\Lambda(z))\\
       \dot{\rho}+3H(\rho+P/c)=0 
\label{Eq:friedmann}
\end{cases}
\end{equation}
with the assumption of zero curvature and the density parameter with two components of matter and dark energy ($\Omega_M+\Omega_{\Lambda}=1$), the Hubble parameter is found as

\begin{equation}
    H(z)=H_0\sqrt{\Omega_M (1+z)^3+\Omega_{\Lambda} e^{3 \int_0 ^z \frac{dz'(1+\omega(z'))}{1+z'}}}
    \label{Eq:Hubble}
\end{equation}
where the $\omega(z)$ is the coefficient of the dark energy equation of state $P= c\omega(z) \rho$. 
From equation \ref{Eq:Hubble}, we can substitute the Hubble parameter.  So the comoving distance of Eq.(\ref{Eq:ComovingDist}) can be written as
\begin{equation}
    D_{C} (z)=\frac{c}{H_0}\int_0 ^z \frac{dz'}{\sqrt{\Omega_M (1+z')^3+\Omega_{\Lambda} e^{3 \int_0 ^{z'} \frac{dz^{''}(1+\omega(z^{''}))}{1+z^{''}}}}}
    \label{Eq:ComovingDistFinal}
\end{equation}
Obviously, $D_{C} (z)$ depends on the $\alpha$ parameter in the dark energy equation of state $\omega(z)=-1+\alpha z$. Difference between $ D_{C} (z)$ of $\omega CDM$ model with $\alpha=0.9$
and $\Lambda CDM$ model is shown in Fig.\ref{Fig_D_C}

By substitution Eq.\ref{Eq:R_Wanderman}, Eq.\ref{Eq:Hubble} and Eq.\ref{Eq:ComovingDistFinal} in Eq.\ref{Eq:Gamma_Def} we can determined theoretical value for $\Gamma(z)$.
 We anticipate that the observed number of GRB events per redshift,$\Gamma_{Obs}(z)$, is obtained from the theoretical $\Gamma(z)$ as

\begin{equation}
    \Gamma_{Obs}(z)=\epsilon(z)\ \Gamma(z) \ ,
    \label{Eq:Gamma_effi_df}
\end{equation}
where $\epsilon(z)$ is the Swift/BAT detection rate function. In the next section, we focus on calculating the detection rate function of Swift/BAT for the $\omega CDM$ model.

\section{Efficiency and observational rate of long GRBs}
\label{Sec:Efficiency}

\begin{figure}[!htbp]
\centering
\includegraphics[width=0.7\linewidth]{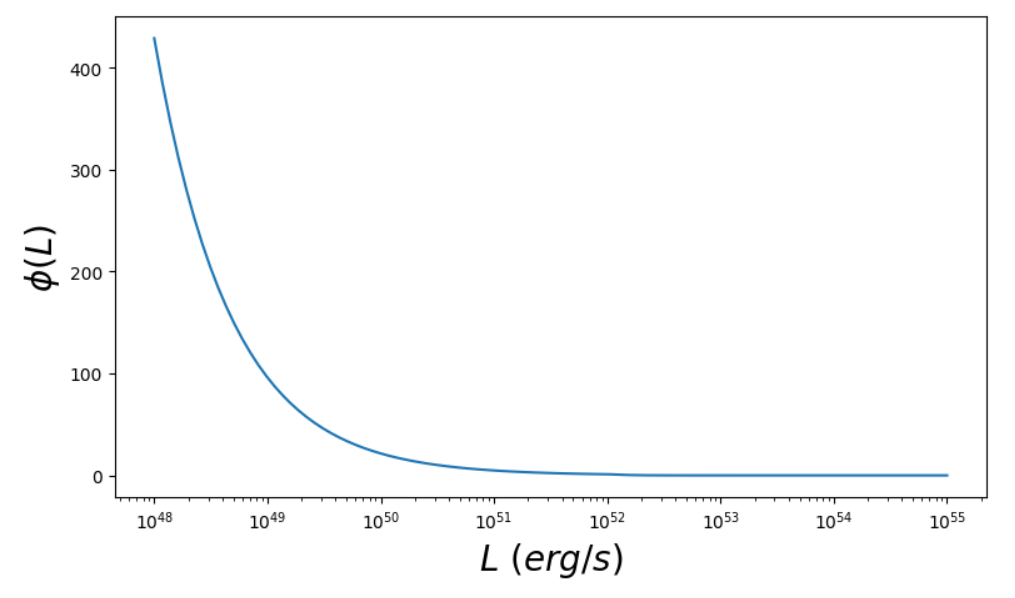}
\caption{\label{Fig_phi} 
Schematic of Eq.(\ref{Eq:phi}) that shows GRBs luminosity distribution function $\phi(L)= \frac{dN}{dL}$
}
\end{figure}

The detection rate function of Swift/BAT, $\epsilon(z)$, can be defined based on BAT efficiency in the detection of peak flux, $\eta(P)$, and GRBs luminosity distribution function, $\phi(L)$, as
\begin{equation}
 \epsilon(z)= \frac{\Delta \Omega}{4 \pi} \ f\eta(P)\ \phi(L)
    \label{Eq:effi}
\end{equation}
where $\Delta \Omega$ is the Swift field of view $(\approx 2 sr)$ \citep{Barthelmy2005}, $f$ is time fraction which BAT dedicated to observing GRBs $(\approx90\%)$\citep{Lien2014}. 
we use luminosity distribution function $\phi(L)$ from 
\cite{Wanderman_2010}
research
\begin{equation}
 \phi(L)= \frac{dN}{dL}=
\begin{cases} 
      (\frac{L}{L_{*}})^x & L < L_{*} \\
      (\frac{L}{L_{*}})^y & L_{*} < L
\label{Eq:phi}
\end{cases}
\end{equation}
where $L$ is the peak luminosity (not the average luminosity), $L_{*}= 10^{52.05} erg/s$, $x= -0.65$, and $y=-3.00$ are determined from results of the best-fit sample with GRB luminosity evolution \citep{Lien2014} (see Fig.\ref{Fig_phi}). 

\begin{figure}[!htbp]
\centering
\includegraphics[width=0.7\linewidth]{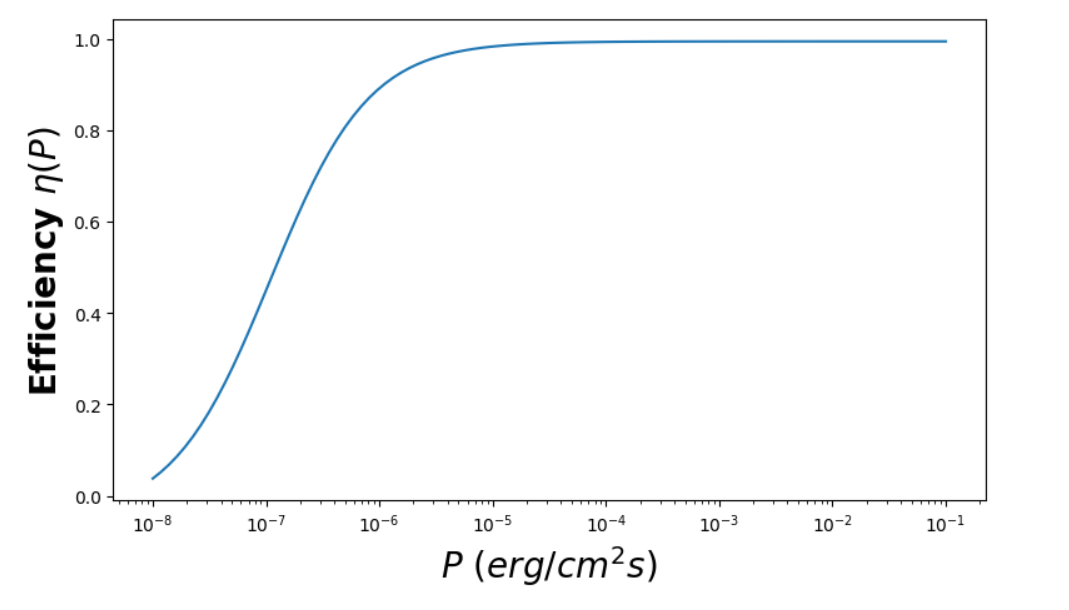}
\caption{\label{Fig_eta} 
BAT triggering efficiency function $\eta(P)$ based on Eq.(\ref{Eq:etta})
}
\end{figure}

On the other hand, \citep{Howell2014} found the BAT triggering efficiency function $(\eta(P))$ based on the comprehensive simulation of peak flux and triggered distribution \citep{Lien2014} study as
\begin{equation}
 \eta(P) = \frac{a(b+c\frac{P}{P_0})}{a+ \frac{1}{d}\frac{P}{ P_0}}
    \label{Eq:etta}
\end{equation}
where $a = 0.47$, $b = -0.05$, $c = 1.46$, $d = 1.45$, and $P_0 = 1.6 \times 10^{-7} erg \ cm^{-2} \ s^{-1}$ (see Fig.\ref{Fig_eta}.). $P$ in the above equation is the average energy flux received by the detector of BAT and therefore $\eta(P)$ is independent of the cosmological model. The average energy flux $P$ in the total energy band is defined as 
\begin{equation}
 P = \frac{L}{4 \pi D_L(z)^2} = \frac{\mathlarger{\int}_{\scriptscriptstyle{ 0}}^{\scriptscriptstyle{\infty}}E\ S(E) dE}{4 \pi D_L(z)^2}
    \label{Eq:P_E}
\end{equation}

where $S(E)$ is the rest frame photon spectrum and $D_L(z)$ is luminosity distance $D_L(z)= (1+z)D_C(z)$.

So, Swift/BAT observed numbers of GRBs per redshift for the general $\omega CDM$ model where the $\omega(z)$ is the coefficient of the dark energy equation of state ($P= \omega(z) \rho$) can be written by substitution of Eq.(\ref{Eq:effi}) in Eq.(\ref{Eq:Gamma_effi_df})
\begin{equation}
    \Gamma_{Obs}(z)=\frac{ \Delta \Omega }{4 \pi} \ f \int_0^{z} \int_{L_{min}(z)}^{L_{max}} \eta(P)\ \phi(L)\ dL\ \Gamma(z) \ dz 
    \label{Eq:Gamma_effi_final}
\end{equation}
where $\Gamma(z)$ determined in Eq.(\ref{Eq:Gamma_Def})

where $P= \frac{L}{4\pi D_L(z)^2}$ and $L_{min}= 4 \pi D_L(z)^2 f_{min}$. If we consider the inclination of GRBs jet toward the observer, $f_{min}= 10^{-7} erg\ cm^{-2} s^{-1}$for extremely off-axis GRBs.

In the following section for numerical calculation of the observational rate of long GRBs, we use Eq.(\ref{Eq:Gamma_Def}) and Eq.(\ref{Eq:Gamma_effi_final}) to
find LGRBs duration histogram of $\Lambda CDM$ model and the cosmological model with dark energy model of $\omega(z)=-1+\alpha z$.

\section{Dependence of long GRBs duration histogram on the cosmological model}
\label{Sec:GRB_Histogram}
\begin{figure}[!htbp]
\centering
\includegraphics[width=0.7\linewidth]{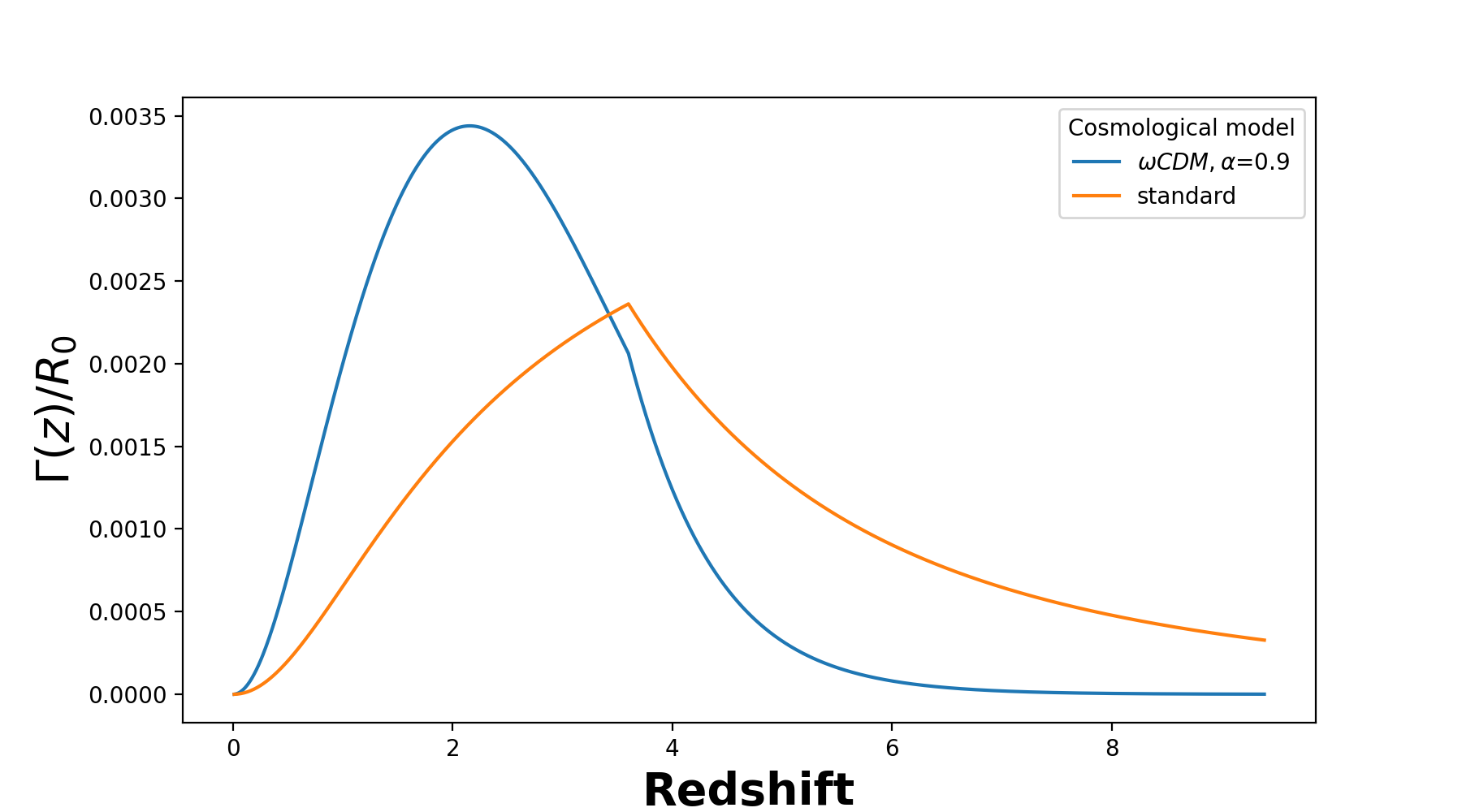}
\caption{\label{fig:Gamma} The orange line shows the theoretical curves of $\Gamma(z)/R_0$ of Eq.(\ref{Eq:Gamma_Def}) for $\Lambda CDM$ model and the blue line depicts this parameter for a cosmological model with dark energy model of $\omega(z)=-1+0.9 z$. Both $\Gamma(z)/R_0$ are normalized to one.  }
\end{figure}

\begin{figure}[!htbp]
\centering
\includegraphics[width=0.7\linewidth]{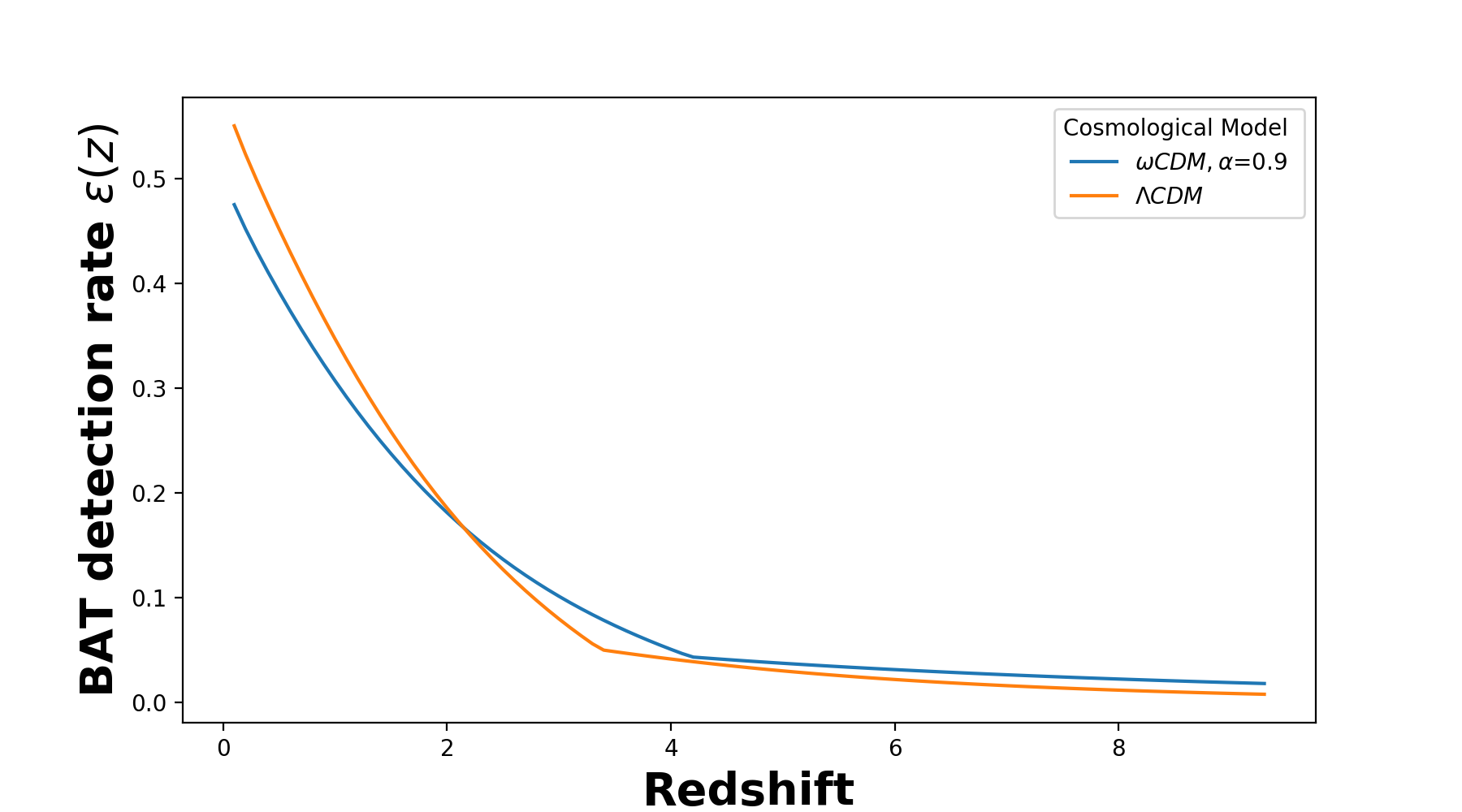}
\caption{\label{fig:efficiency_BAT} Swift/BAT detection rate of long GRBs for $\Lambda CDM$ model and blue line depicted this parameter for a cosmological model with dark energy model of $\omega(z)=-1+0.9 z$. }
\end{figure}

\begin{figure}[!htbp]
\centering
\includegraphics[width=0.7\linewidth]{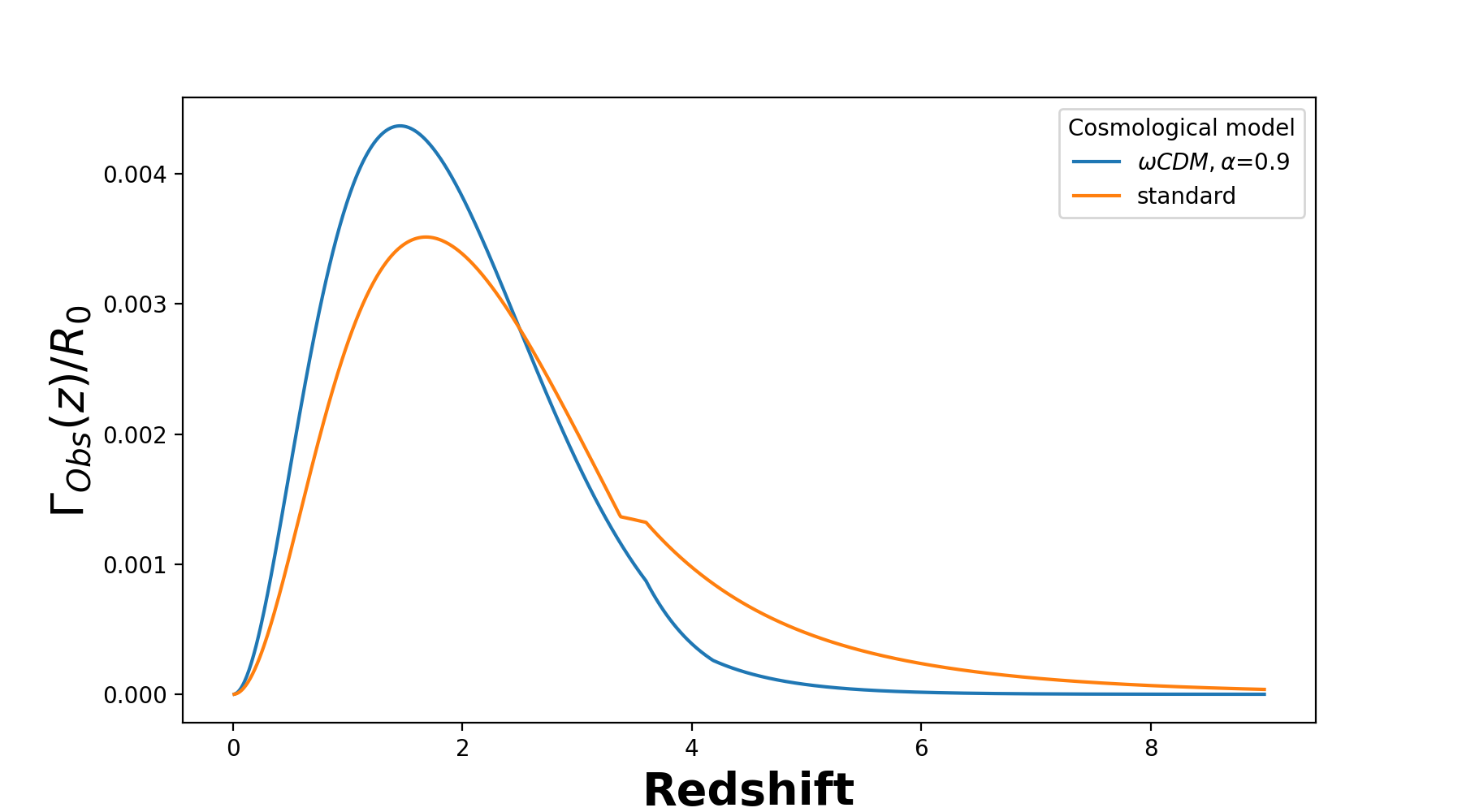}
\caption{\label{fig:Gamma_effi} The orange line shows the  normalized $\Gamma_{Obs}(z)/R_0$ for $\Lambda CDM$ model and the blue line depicts this parameter for a cosmological model with dark energy model of $\omega(z)=-1+0.9 z$}
\end{figure}

\begin{figure}[!htbp]
\centering
\includegraphics[width=0.7\linewidth]{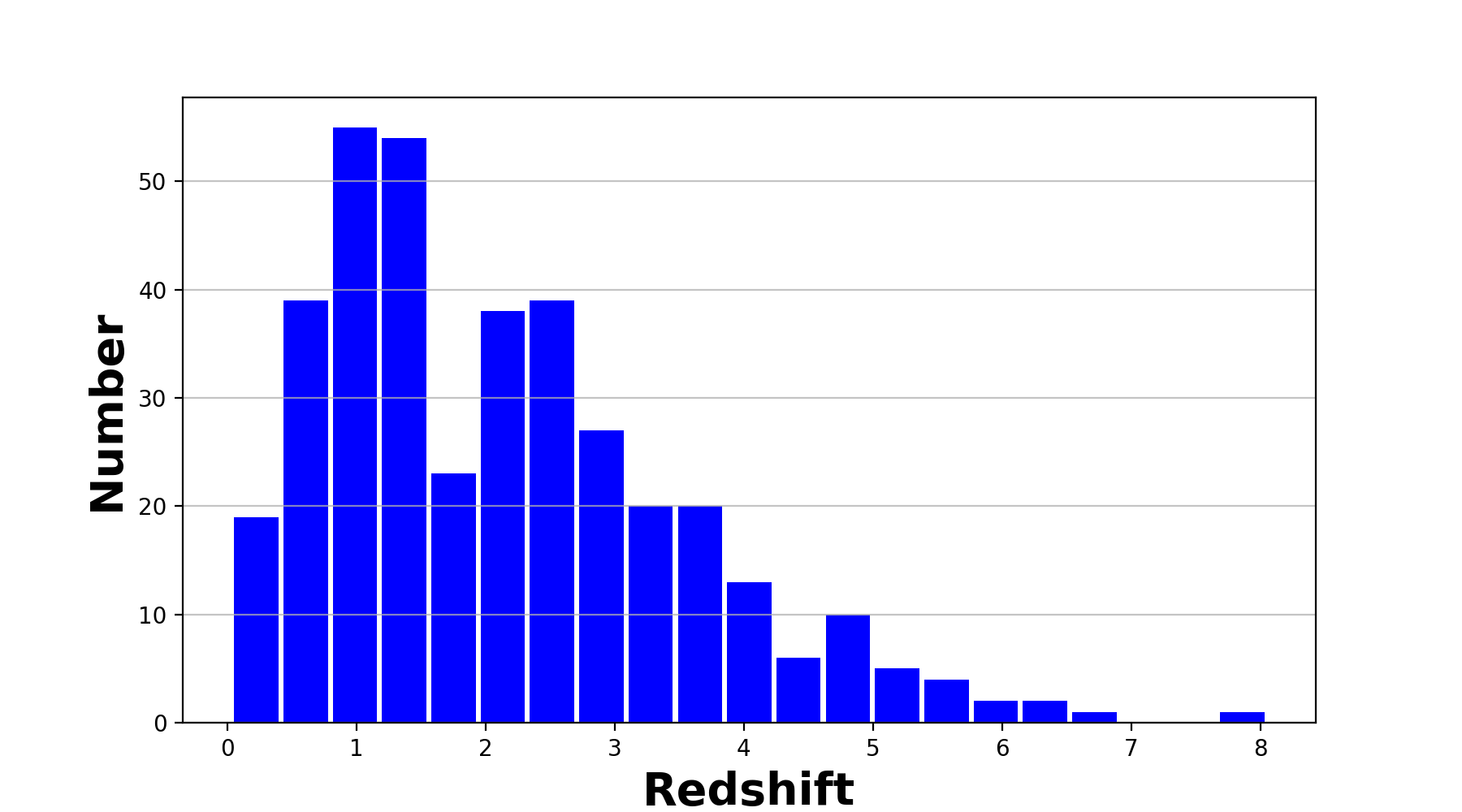}
\caption{\label{fig:z_original} The histogram of the redshift of 385 LGRBs detected by Swift (data from the Swift archive.)}
\end{figure}

In this section, we consider two cosmological models for $\omega(z)$ first $\Lambda CDM$ model with $\omega(z)=-1$ and second a model with $\omega(z)=-1 +\alpha z$, where we assume $\alpha=0.9$.
we want to compare the histogram of LGRBs duration resulting from this $\omega CDM$ model and $\Lambda CDM$ model. First, from Eq.(\ref{Eq:Gamma_Def}) we calculate $\Gamma(z)$ for $\Lambda CDM$ and $\omega CDM$ with $\omega(z)=-1 +0.9 z$ models. The results for $\Gamma(z)/R_0$ are shown in Fig.\ref{fig:Gamma}.
 Then the detection rate per redshift for two cosmological models ($\Lambda CDM$ and $\omega CDM$ with $\omega(z)=-1 +0.9 z$) are calculated by Eq.(\ref{Eq:effi}) that are shown in Fig.\ref{fig:efficiency_BAT}.
From EQ.(\ref{Eq:Gamma_effi_final}) we find normalized $\Gamma_{Obs}(z)$ as $\int _0 ^{z_{max}} \Gamma_{Obs}(z) = 1$ where $z_{max}$ is the maximum redshift of observed GRBs is taken place at $z_{max} = 9.4$. The normalized $\Gamma_{Obs}(z)$ for two cosmological models are depicted in Fig.\ref{fig:Gamma_effi}. The cosmological redshift of 385 Swift LGRBs was reported in Swift archive\footnote{\href{https://swift.gsfc.nasa.gov/archive/grb_table/} {https://swift.gsfc.nasa.gov/archive/grb\_table/}} (up to 23/04/2024). We showed the histogram of the redshift of these LGRBs in Fig.\ref{fig:z_original} that has a shape similar to $\Gamma_{Obs}(z)$ in Fig.\ref{fig:Gamma_effi}.
We use these probability density functions for generating the redshift of LGRBs in our Monte-Carlo simulation.

\begin{figure}[!htbp]
\centering
\includegraphics[width=0.7\linewidth]{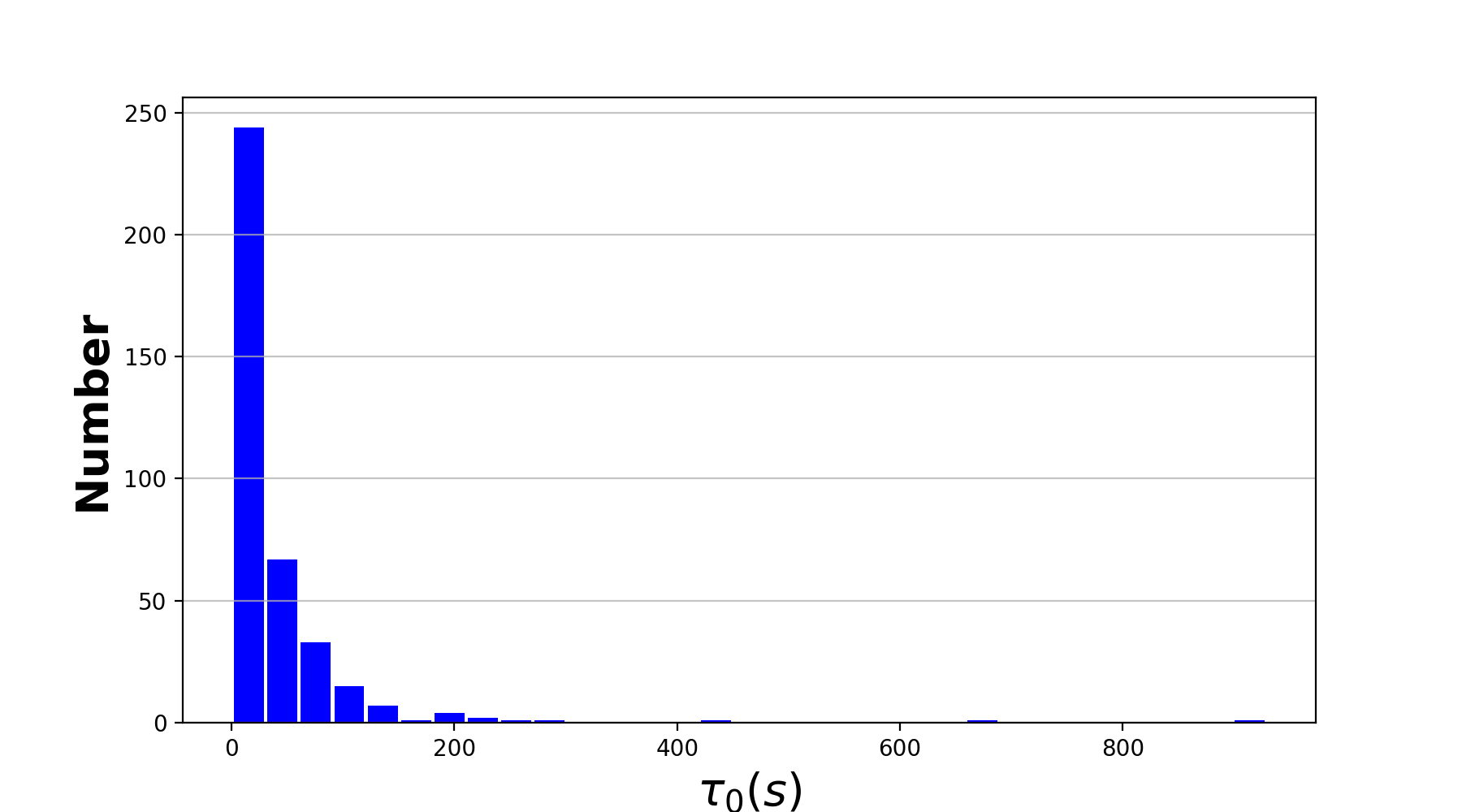}
\caption{\label{fig:tau0_original} The histogram of duration ($T_{90}$) of long GRBs in its comoving frame that is derived from redshifts and duration of these 385 LGRBs detected by Swift (data from the Swift archive).}
\end{figure}

\begin{figure}[!htbp]
\centering
\includegraphics[width=0.7\linewidth]{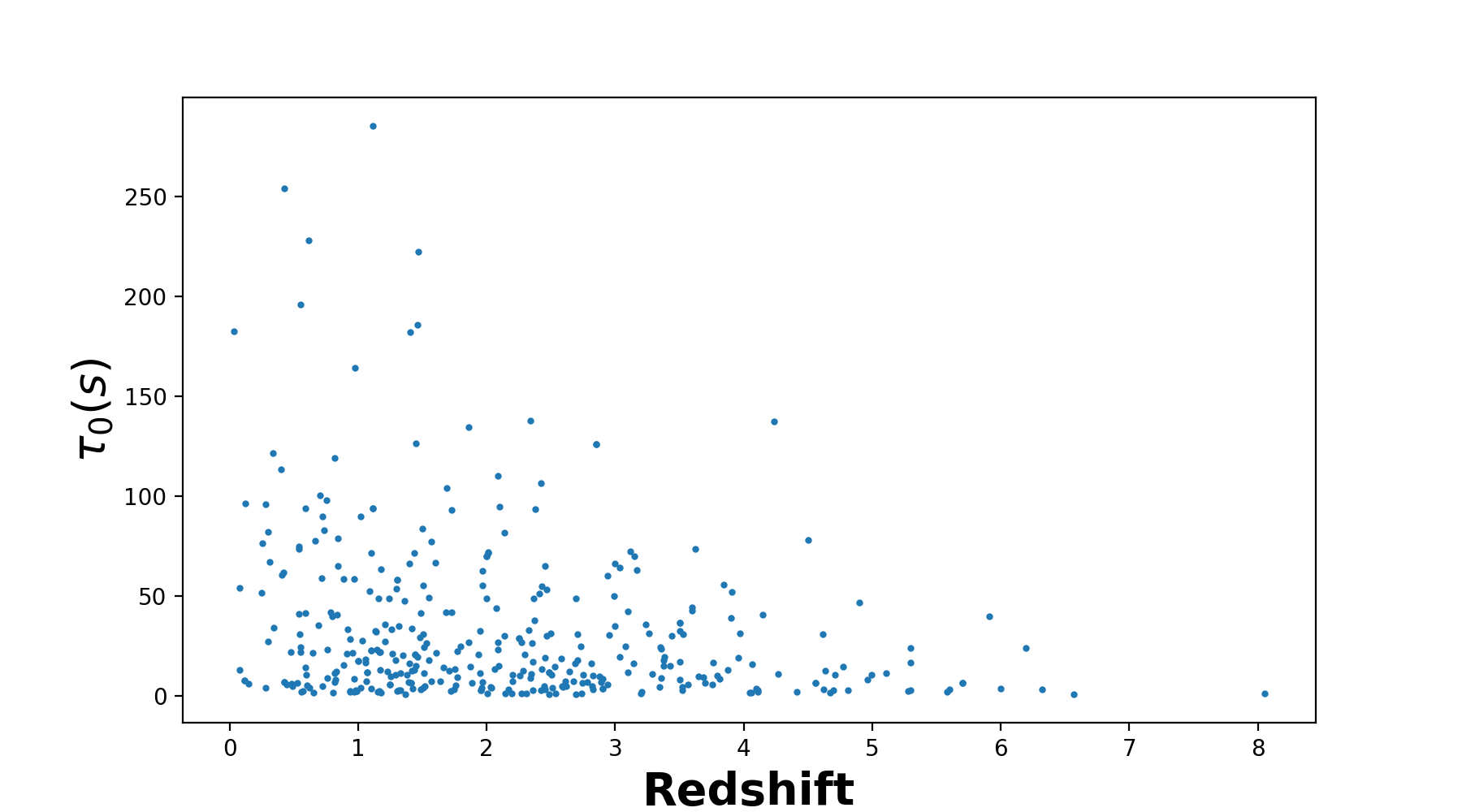}
\caption{\label{fig:ZandTau} $\tau_0$ is the duration of burst ($T_{90}/(1+z)$) in its rest frame which is plotted versus burst redshift. (data from the Swift archive).)}
\end{figure}

For the second step of Monte-Carlo simulation, we need a histogram of LGRBs duration in their comoving frame. The cosmological redshift and $T_{90}$ of 385 Swift LGRBs were reported in this archive.  If $\tau$ is taken as the observed duration of burst ($\tau=T_{90}$), we can detect the duration of the burst at the source frame ($\tau_0$), as $\tau_0=\tau/(1+z)$. We find the histogram of  $\tau_0$ of these 385 LGRBs that is shown in Fig.\ref{fig:tau0_original}. Two parameters of $\tau_0=\tau/(1+z)$ and redshift are independent (see Fig \ref{fig:ZandTau} and \citep{Horvath2022,Lien3thCatalog}). We examine this point from the normalized correlation coefficient definition,
\begin{equation}
c  = \frac{\sum _{1}^{385} \tau_0(i)z(i) }{\sum_{1}^{385} (\tau_0(i)-\Bar{\tau_0})^2 \sum_{1}^{385}  (z(i)-\Bar{z})^2}
    \label{Eq:corr}
\end{equation}
we find $c=-0.09$ for redshift and the burst at the source frame, which means $\tau_0$ and redshift are almost uncorrelated.
Therefore, since $\tau_0=\tau/(1+z)$ and redshift are uncorrelated, we used $\tau_0$ histogram of 385 LGRBs to generate more $\tau_{0}$ for mock LGRBs in the Monte-Carlo simulation.  


\begin{figure}[!htbp]
\centering
\includegraphics[width=0.9\linewidth]{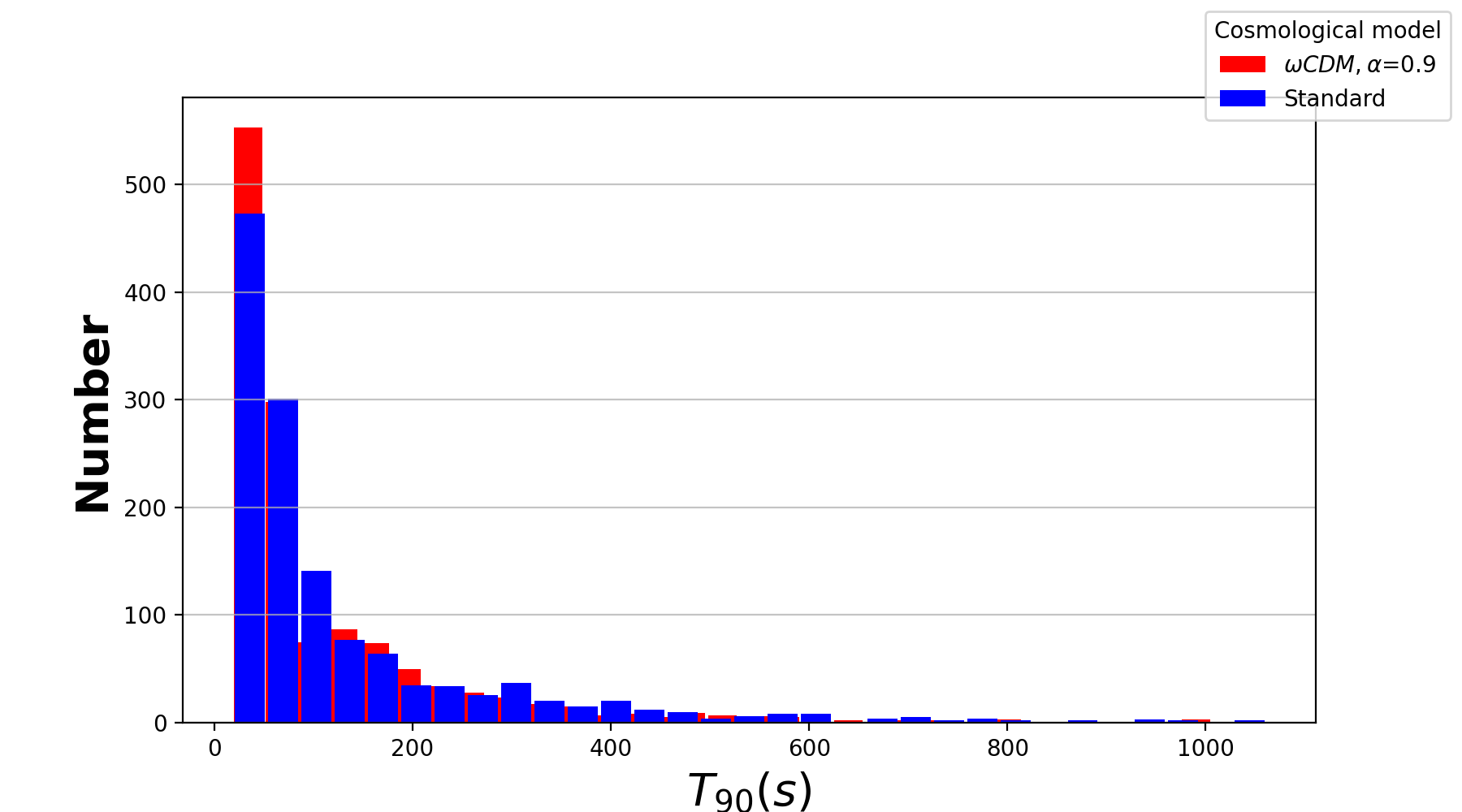}
\caption{Histogram of $\tau=\tau_{0}(1+z)$ regarding the efficiency in detection of Swift/BAT for 1347 mock events and $\Lambda CDM$ model (blue) and the cosmological model with dark energy model of $\omega(z)=-1+0.9 z$ (red).}
\label{fig:histogram_tau_efficency_used}
\end{figure}

\begin{figure}[!htbp]
\centering
\includegraphics[width=0.7\linewidth]{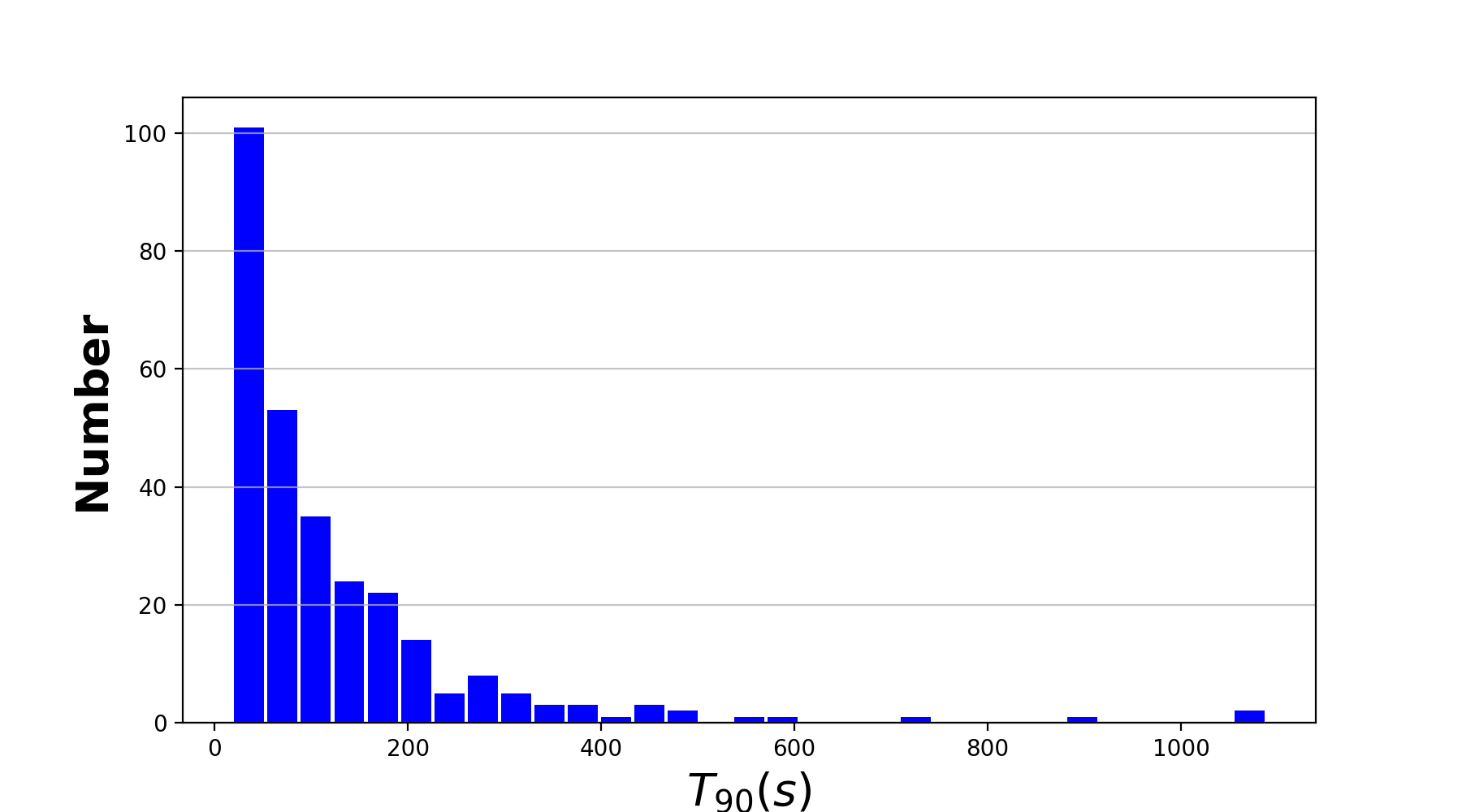}
\caption{\label{fig:T_long_BAT} The histogram of duration ($T_{90}$) of 1347 Swift LGRBs (data from the Swift archive.).}
\end{figure}

Finally, we want to generate the histogram of observed LGRBs duration ($\tau =T_{90}$) for a cosmological model.
From 1389 LGRBs in the Swift archive (up to 23/04/2024), 1347 LGRBs have peak flux $\geq 10^{-7} erg/cm^2s$ and their $T_{90}$ are determined but only redshifts of 385 LGRBs had measured. Therefore, in a Monte Carlo simulation, we want to simulate $\tau = \tau_0 (1+z)$ for 1347 mock LGRBs and compare the histogram of $\tau$ for assumed cosmological model and observational histogram of $\tau$. Because the duration of LGRBs in the Swift archive seems uncorrelated to their redshift, We simulate $\tau = \tau_0 (1+z)$ for 1347 mock LGRBs by using the redshift distribution of Fig.\ref{fig:Gamma_effi} and $\tau_0$ from Fig.\ref{fig:tau0_original} distribution. The histograms of $\tau$ for two cosmological models are depicted in Fig.\ref{fig:histogram_tau_efficency_used}. As we expect, the $\tau$ histogram of LGRBs depends on the cosmological model. The histogram of the $T_{90}$ of real data of Swift LGRBs archive is depicted in Fig.\ref{fig:T_long_BAT}.  In the following section, we compare $\tau$ histogram of LGRBs to find the best $\alpha$ for the cosmological model with a dark energy model of $\omega(z)=-1+\alpha z$ based on $\chi^2$ test on LGRBs duration histogram.


\section{Best Parameter of $\alpha$ for the Cosmological Model}
\label{Sec:w}
\begin{figure}[!htbp]
\centering
\includegraphics[width=0.7\linewidth]{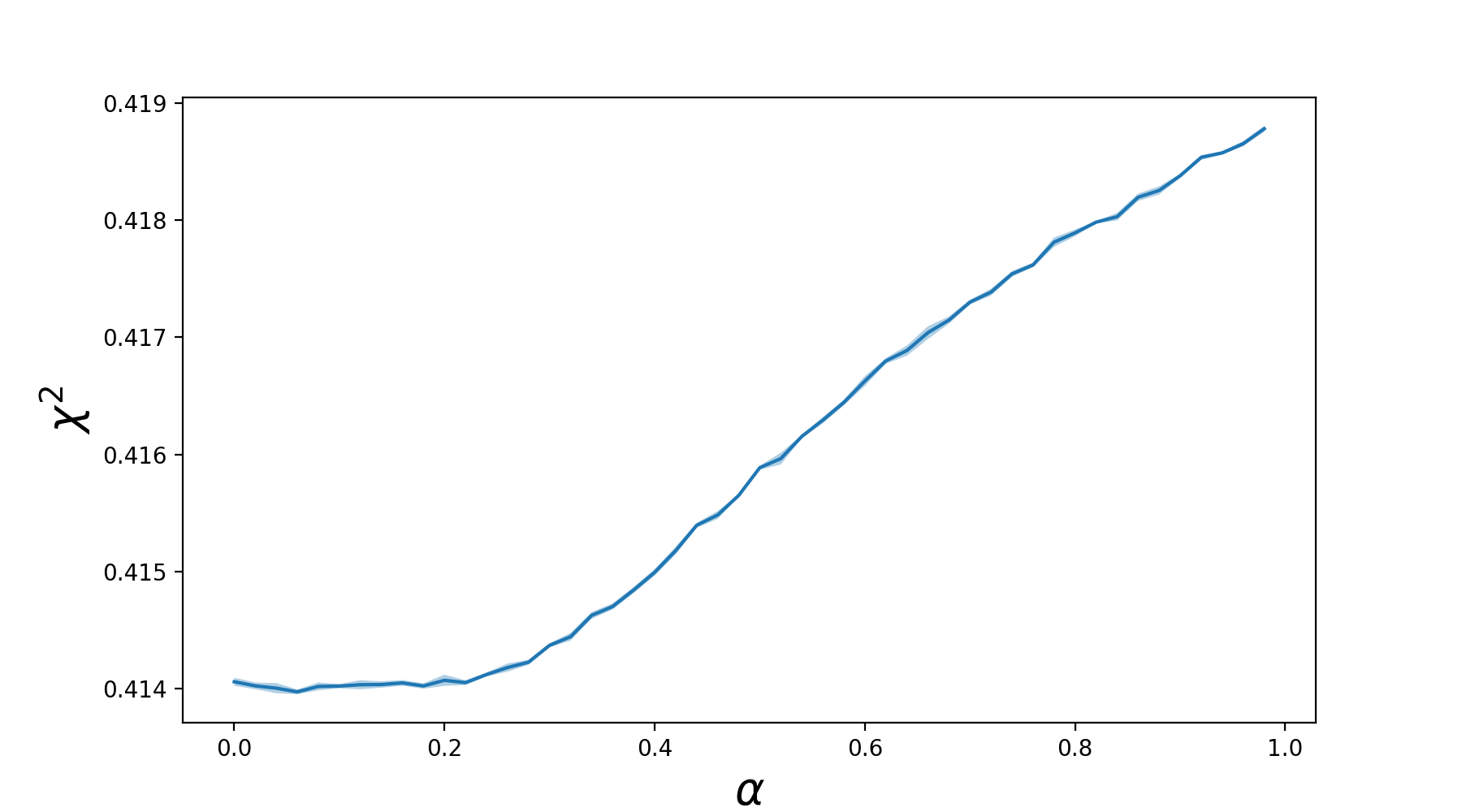}
\caption{\label{fig:W_best} By using Eq.(\ref{Eq:chi}) $\chi^2$ for cosmological model with $\omega(z)= -1 + \alpha z$ is calculated in the range of $\alpha=[0,1]$ an with $\Delta \alpha = 0.02$.  }
\end{figure}

As we interpreted in the previous section, we use the numerical calculation of the observational LGRBs rate to find the LGRBs duration histogram of the cosmological model with a dark energy model of $\omega(z)=-1+\alpha z$. Here we want to find the best parameter of $\alpha$ from the histograms of mock LGRBs duration. We assume the range of $\alpha =[0,1]$ for the $\omega CDM$ cosmological model with $\omega(z)= -1 + \alpha z$ and then find the LGRBs duration ($T_{90}$) histogram. Subsequently, we apply a $\chi^2$ method. The $\chi^2$ of histogram is defined as
\begin{equation}
\chi^2  = \frac{1}{n-1}\sum _i^n \frac{(N_{obs} (i)-N_{theory}(i))^2}{N_{obs} (i) + N_{theory}(i)}
    \label{Eq:chi}
\end{equation}
Where i is each histogram bin and $n$ is the number of data in summation. $N_{theory}$ in the denominator is due to the Poison's noise in measurement. 
We assume the range of $\alpha =[0,1]$ for the cosmological model and then calculate $\chi^2$ for each $\alpha$ in the range of $\alpha =[0,1]$. To find the error in $\chi^2$, for all $\alpha =[0,1]$ we run 10 times the code. In Fig. \ref{fig:W_best} the mean of  $\chi^2$ for each  $\alpha$ depicted by the line and the variances of $\chi^2$ are shown by the shaded region. This figure shows $\chi^2$ has a local minimum at $\alpha=0.08 ^{+ 0.04}_{-0.02}$. 
Therefore, the $\omega CDM$ depicts more consistency with the histogram of long GRBs duration than the standard $\Lambda CDM$ cosmological model

%

\begin{figure}[!htbp]
\centering
\includegraphics[width=0.7\linewidth]{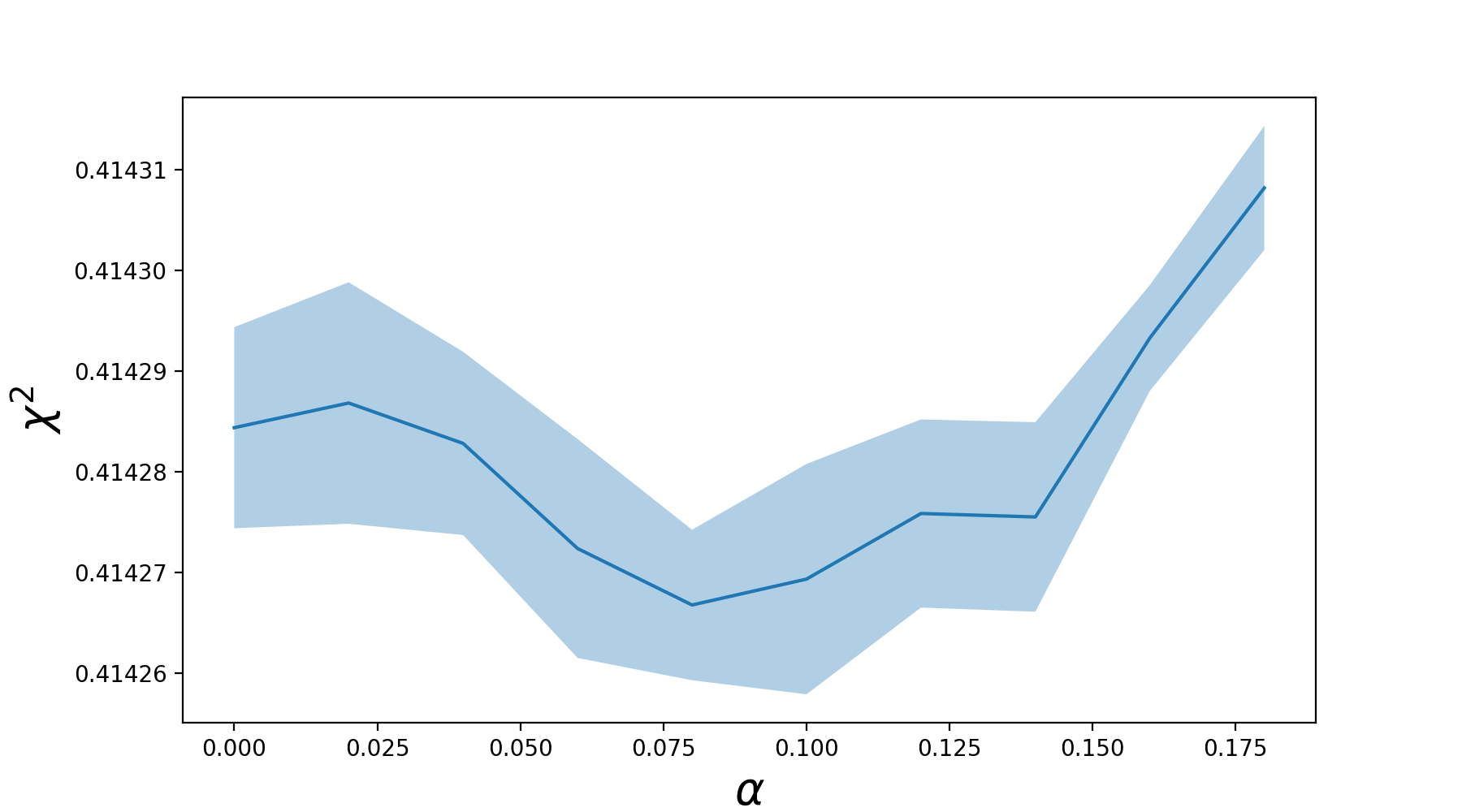}
\caption{\label{fig:W_best_zoomed} The $\chi^2$ for cosmological model with $\omega(z)= -1 + \alpha z$ around the minimum value of $\alpha=0.08 ^{+ 0.04}_{-0.02}$  }
\end{figure}

\section{Conclusion}
\label{Sec:Conclusion}
Gamma-ray bursts, acting as powerful signals in the depths of the Universe. 
The Swift satellite has measured the duration of approximately one thousand gamma-ray bursts. The idea of this paper is to use the long gamma-ray bursts duration histogram to constrain the values of $\omega$ in the $\omega CDM$ cosmology model.
One clear theoretical explanation for the accelerating universe is the presence of a cosmological constant.
The $\Lambda CDM$ model cosmology presents a straightforward and coherent model that has agreed with observations \citep{1998Riess,1999Perlmutter}. Nevertheless, recent observations have indicated an increasing cosmic tension in the measurement of the Hubble parameter \citep{2022Riess}. $H_0$ tension problem led to searching for alternative cosmological models. Such as $\omega CDM$ models propose that dark energy equations ($P= c \omega(z) \rho$) can change over time with redshift meaning that cosmological constant of $\Lambda CDM$ replaces with an evolving dark energy component ($\omega CDM$),
which caused modifications of standard gravity ($f(R)$-gravity). Such modified gravity models suggest that the acceleration of cosmic expansion may be driven by a different form of energy like a scalar field \citep{Baghram2007}.

In this project, we introduce the $\chi^2$ method to compare the histogram of LGRBs duration based on different cosmological models by using a dataset of long GRBs from the Swift/BAT archive.
We compared $ \omega CDM$ model 
with the dark energy equation of state, $P= c \omega(z) \rho$, where $\omega(z)$ the form of $1+\alpha z$ and the $\Lambda CDM$ model with observation data of long GRBs duration.
In our Monte-Carlo simulation, We generate $T_{90} = \tau_0 (1+z)$ for 1347 simulated LGRBs. The redshift distribution is derived from calculating the observational rate of LGRBs per redshift, while the distribution is based on 385 LGRBs whose redshifts were determined by Swift.
Then, the $\chi^2$ minimization technique was applied to find the optimal value of  $\alpha$ in the dark energy equation of state, with $\omega(z)=1+\alpha z$. We found that if $\alpha=0.08 ^{+ 0.04}_{-0.02}$, so histogram of LGRBs duration is best fitted to real data. Therefore, the analysis showed the $\omega CDM$ model is slightly preferred over the $\Lambda CDM$ model.
The values of $\alpha$ in the dark energy equation of state from the SNIa data by Pantheon and DES-SN3YR samples and from the combinations of SNIa, CMB, and BAO  data in other study show that  $\alpha $ is not zero and equals to $ −0.129 \pm 0.755$  for Pantheon-SN+CMB \citep{2016PlanckCollaboration, Mazumdar2021}, $\alpha = −0.126 \pm 0.384$ for Pantheon-SN+CMB+BAO \citep{2011Beutler, 2014Anderson, 2015Ross, 2016PlanckCollaboration, Mazumdar2021}, and $\alpha =  −0.387 \pm 0.430 $ for  DES-SN3YR+CMB+BAO data \citep{Dark_energy_Abbott, Mazumdar2021}.


\section*{ACKNOWLEDGEMENT}
This work was supported by Iran National Elites Foundation grant\\
\bibliographystyle{aasjournal}
\bibliography{ref}

\begin{thebibliography}{}
\expandafter\ifx\csname natexlab\endcsname\relax\def\natexlab#1{#1}\fi
\providecommand{\url}[1]{\href{#1}{#1}}
\providecommand{\dodoi}[1]{doi:~\href{http://doi.org/#1}{\nolinkurl{#1}}}
\providecommand{\doeprint}[1]{\href{http://ascl.net/#1}{\nolinkurl{http://ascl.net/#1}}}
\providecommand{\doarXiv}[1]{\href{https://arxiv.org/abs/#1}{\nolinkurl{https://arxiv.org/abs/#1}}}

\bibitem[{{Abbott} {et~al.}(2019){Abbott}, {Allam}, {Andersen}, {Angus}, \& et~al.}]{Dark_energy_Abbott}
{Abbott}, T., {Allam}, S., {Andersen}, P., {Angus}, \& et~al. 2019, apjl, 872, L30

\bibitem[{{Alam} {et~al.}(2017){Alam}, {Ata}, {Bailey}, \& et~al.}]{Alam2017}
{Alam}, S., {Ata}, M., {Bailey}, \& et~al. 2017, mnras, 470, 2617

\bibitem[{{Amati} {et~al.}(2002){Amati}, {Frontera}, {Tavani}, {in t Zand}, {Antonelli}, {Costa}, {Feroci}, {Guidorzi}, {Heise}, {Masetti}, {Montanari}, {Nicastro}, {Palazzi}, {Pian}, {Piro}, \& {Soffitta}}]{Amati}
{Amati}, L., {Frontera}, F., {Tavani}, M., {et~al.} 2002, aap, 390, 81

\bibitem[{{Anderson} {et~al.}(2014){Anderson}, {Aubourg}, {Bailey}, \& et~al}]{2014Anderson}
{Anderson}, L., {Aubourg}, E., {Bailey}, \& et~al. 2014, mnras, 441, 24

\bibitem[{{Barthelmy} {et~al.}(2005){Barthelmy}, {Barbier}, {Cummings}, \& et~al}]{Barthelmy2005}
{Barthelmy}, S.~D., {Barbier}, L.~M., {Cummings}, \& et~al. 2005, ssr, 120, 143

\bibitem[{{Berger}(2014)}]{Berger}
{Berger}, E. 2014, araa, 52, 43

\bibitem[{{Beutler} {et~al.}(2011){Beutler}, {Blake}, {Colless}, {Jones}, {Staveley-Smith}, {Campbell}, {Parker}, {Saunders}, \& {Watson}}]{2011Beutler}
{Beutler}, F., {Blake}, C., {Colless}, M., {et~al.} 2011, mnras, 416, 3017

\bibitem[{{Cucchiara} {et~al.}(2011){Cucchiara}, {Levan}, {Fox}, \& {Tanvir}}]{Cucchiara}
{Cucchiara}, A., {Levan}, A., {Fox}, D.~B., \& {Tanvir}, a. e.~a. 2011, apj, 736, 7

\bibitem[{{Dai} {et~al.}(2021){Dai}, {Zheng}, {Li}, {Gao}, \& {Zhu}}]{Dai}
{Dai}, Y., {Zheng}, X.-G., {Li}, Z.-X., {Gao}, H., \& {Zhu}, Z.-H. 2021, aap, 651

\bibitem[{{Dainotti} {et~al.}(2008){Dainotti}, {Cardone}, \& {Capozziello}}]{Dainotti}
{Dainotti}, M.~G., {Cardone}, V.~F., \& {Capozziello}, S. 2008, mnras, 391, L79

\bibitem[{{Dainotti} {et~al.}(2022{\natexlab{a}}){Dainotti}, {Nielson}, {Sarracino}, {Rinaldi}, {Nagataki}, {Capozziello}, {Gnedin}, \& {Bargiacchi}}]{Dainotti2}
{Dainotti}, M.~G., {Nielson}, V., {Sarracino}, G., {et~al.} 2022{\natexlab{a}}, mnras, 514, 1828

\bibitem[{{Dainotti} {et~al.}(2022{\natexlab{b}}){Dainotti}, {Sarracino}, \& {Capozziello}}]{Dainotti3}
{Dainotti}, M.~G., {Sarracino}, G., \& {Capozziello}, S. 2022{\natexlab{b}}, pasj, 74, 1095

\bibitem[{{Demianski} {et~al.}(2017){Demianski}, {Piedipalumbo}, {Sawant}, \& {Amati}}]{Demianski}
{Demianski}, M., {Piedipalumbo}, E., {Sawant}, D., \& {Amati}, L. 2017, aap, 598, A112

\bibitem[{{Fenimore} \& {Ramirez-Ruiz}(2000)}]{Fenimore}
{Fenimore}, E.~E., \& {Ramirez-Ruiz}, E. 2000, arXiv e-prints, astro

\bibitem[{{Firmani} {et~al.}(2006){Firmani}, {Ghisellini}, {Avila-Reese}, \& {Ghirlanda}}]{Firmani}
{Firmani}, C., {Ghisellini}, G., {Avila-Reese}, V., \& {Ghirlanda}, G. 2006, mnras, 370, 185

\bibitem[{{Fryer} {et~al.}(2007){Fryer}, {Hungerford}, \& {Young}}]{Fryer}
{Fryer}, C.~L., {Hungerford}, A.~L., \& {Young}, P.~A. 2007, apjl, 662, L55

\bibitem[{{Ghirlanda} {et~al.}(2006){Ghirlanda}, {Ghisellini}, \& {Firmani}}]{Ghirlanda2}
{Ghirlanda}, G., {Ghisellini}, G., \& {Firmani}, C. 2006, New Journal of Physics, 8, 123

\bibitem[{{Ghirlanda} {et~al.}(2004){Ghirlanda}, {Ghisellini}, \& {Lazzati}}]{Ghirlanda}
{Ghirlanda}, G., {Ghisellini}, G., \& {Lazzati}, D. 2004, apj, 616, 331

\bibitem[{{Hjorth} {et~al.}(2003){Hjorth}, {Sollerman}, {M{o}ller}, \& et~al}]{Hjorth2003}
{Hjorth}, J., {Sollerman}, J., {M{o}ller}, \& et~al. 2003, nat, 423, 847

\bibitem[{{Horvath} {et~al.}(2022){Horvath}, {Racz}, {Bagoly}, {Balazs}, \& {Pinter}}]{Horvath2022}
{Horvath}, I., {Racz}, I.~I., {Bagoly}, Z., {Balazs}, L.~G., \& {Pinter}, S. 2022, Universe, 8, 221

\bibitem[{{Howell} {et~al.}(2014){Howell}, {Coward}, {Stratta}, {Gendre}, \& {Zhou}}]{Howell2014}
{Howell}, E., {Coward}, D., {Stratta}, G., {Gendre}, B., \& {Zhou}, H. 2014, mnras, 444, 15

\bibitem[{{Hu} {et~al.}(2021){Hu}, {Wang}, \& {Dai}}]{Hu}
{Hu}, J.~P., {Wang}, F.~Y., \& {Dai}, Z.~G. 2021, mnras, 507, 730

\bibitem[{{Izzo} {et~al.}(2015){Izzo}, {Muccino}, {Zaninoni}, {Amati}, \& {Della Valle}}]{Izzo}
{Izzo}, L., {Muccino}, M., {Zaninoni}, E., {Amati}, L., \& {Della Valle}, M. 2015, aap, 582, A115

\bibitem[{{Khadka} {et~al.}(2021){Khadka}, {Luongo}, {Muccino}, \& {Ratra}}]{Khadka}
{Khadka}, N., {Luongo}, O., {Muccino}, M., \& {Ratra}, B. 2021, jcap, 2021, 042

\bibitem[{{Kouveliotou} {et~al.}(1993){Kouveliotou}, {Meegan}, {Fishman}, {Bhat}, {Briggs}, {Koshut}, {Paciesas}, \& {Pendleton}}]{Kouveliotou}
{Kouveliotou}, C., {Meegan}, C.~A., {Fishman}, G.~J., {et~al.} 1993, apjl, 413

\bibitem[{{Kulkarni} {et~al.}(1999){Kulkarni}, {Djorgovski}, {Odewahn}, {Bloom}, {Gal}, {Koresko}, {Harrison}, {Lubin}, {Armus}, {Sari}, {Illingworth}, {Kelson}, {Magee}, {van Dokkum}, {Frail}, {Mulchaey}, {Malkan}, {McClean}, {Teplitz}, {Koerner}, {Kirkpatrick}, {Kobayashi}, {Yadigaroglu}, {Halpern}, {Piran}, {Goodrich}, {Chaffee}, {Feroci}, \& {Costa}}]{Kulkarni}
{Kulkarni}, S.~R., {Djorgovski}, S.~G., {Odewahn}, S.~C., {et~al.} 1999, nat, 398

\bibitem[{{Lan} {et~al.}(2023){Lan}, {Gao}, {Li}, {Xiao}, {Ai}, {Peng}, {Li}, {Wang}, {Xu}, {Lin}, {Lei}, {Zhang}, {Zhang}, {Zheng}, {Liu}, {Xue}, {Wang}, {Tan}, \& {Xiong}}]{Lan2023}
{Lan}, L., {Gao}, H., {Li}, A., {et~al.} 2023, apjl, 949, L4

\bibitem[{{Li} {et~al.}(2023){Li}, {Yang}, {Yi}, {Hu}, {Wang}, \& {Qu}}]{Li}
{Li}, J.-L., {Yang}, Y.-P., {Yi}, S.-X., {et~al.} 2023, apj, 953, 58

\bibitem[{{Liang} \& {Zhang}(2005)}]{Liang}
{Liang}, E., \& {Zhang}, B. 2005, apj, 633, 611

\bibitem[{{Lien} {et~al.}(2014){Lien}, {Sakamoto}, \& {Gehrels}}]{Lien2014}
{Lien}, A., {Sakamoto}, T., \& {Gehrels}, a. e.~a. 2014, apj, 783, 24

\bibitem[{{Lien} {et~al.}(2016){Lien}, {Sakamoto}, {Barthelmy}, {Baumgartner}, {Cannizzo}, {Chen}, {Collins}, {Cummings}, {Gehrels}, {Krimm}, {Markwardt}, {Palmer}, {Stamatikos}, {Troja}, \& {Ukwatta}}]{Lien3thCatalog}
{Lien}, A., {Sakamoto}, T., {Barthelmy}, S.~D., {et~al.} 2016, apj, 829, 7

\bibitem[{{Lin} {et~al.}(2015){Lin}, {Li}, {Wang}, \& {Chang}}]{Lin}
{Lin}, H.-N., {Li}, X., {Wang}, S., \& {Chang}, Z. 2015, mnras, 453, 128

\bibitem[{{Mazumdar} {et~al.}(2021){Mazumdar}, {Mohanty}, \& {Parashari}}]{Mazumdar2021}
{Mazumdar}, A., {Mohanty}, S., \& {Parashari}, P. 2021, European Physical Journal Special Topics, 230, 2055

\bibitem[{{Movahed} {et~al.}(2007){Movahed}, {Baghram}, \& {Rahvar}}]{Baghram2007}
{Movahed}, M.~S., {Baghram}, S., \& {Rahvar}, S. 2007, prd, 76, 044008

\bibitem[{Norris {et~al.}(2000)Norris, Marani, \& Bonnell}]{Norris}
Norris, J.~P., Marani, G.~F., \& Bonnell, J.~T. 2000, The Astrophysical Journal, 534, 248, \dodoi{10.1086/308725}

\bibitem[{{Perlmutter} {et~al.}(1998){Perlmutter}, {Aldering}, {della Valle}, \& et~al}]{Perlmutter}
{Perlmutter}, S., {Aldering}, G., {della Valle}, \& et~al. 1998, nat, 391, 51

\bibitem[{{Perlmutter} {et~al.}(1999){Perlmutter}, {Aldering}, {Goldhaber}, \& et~al}]{1999Perlmutter}
{Perlmutter}, S., {Aldering}, G., {Goldhaber}, \& et~al. 1999, apj, 517, 565

\bibitem[{{Planck Collaboration} {et~al.}(2016){Planck Collaboration}, {Ade}, {Aghanim}, {Arnaud}, \& {Ashdown}}]{2016PlanckCollaboration}
{Planck Collaboration}, {Ade}, P.~A.~R., {Aghanim}, N., {Arnaud}, M., \& {Ashdown}, a. e.~a. 2016, aap, 594, A13

\bibitem[{{Riess} {et~al.}(2022){Riess}, {Yuan}, {Macri}, \& et~al}]{2022Riess}
{Riess}, A.~G., {Yuan}, W., {Macri}, \& et~al. 2022, apjl, 934, L7

\bibitem[{{Riess} {et~al.}(1998){Riess}, {Filippenko}, {Challis}, {Clocchiatti}, {Diercks}, {Garnavich}, {Gilliland}, {Hogan}, {Jha}, {Kirshner}, {Leibundgut}, {Phillips}, {Reiss}, {Schmidt}, {Schommer}, {Smith}, {Spyromilio}, {Stubbs}, {Suntzeff}, \& {Tonry}}]{1998Riess}
{Riess}, A.~G., {Filippenko}, A.~V., {Challis}, P., {et~al.} 1998, aj, 116, 1009

\bibitem[{{Ross} {et~al.}(2015){Ross}, {Samushia}, {Howlett}, {Percival}, {Burden}, \& {Manera}}]{2015Ross}
{Ross}, A.~J., {Samushia}, L., {Howlett}, C., {et~al.} 2015, mnras, 449, 835

\bibitem[{{Schaefer}(2007)}]{Schaefer}
{Schaefer}, B.~E. 2007, apj, 660, 16

\bibitem[{{Tsutsui} {et~al.}(2009){Tsutsui}, {Nakamura}, {Yonetoku}, {Murakami}, {Kodama}, \& {Takahashi}}]{Tsutsui}
{Tsutsui}, R., {Nakamura}, T., {Yonetoku}, D., {et~al.} 2009, jcap, 2009, 015

\bibitem[{{Wanderman} \& {Piran}(2010)}]{Wanderman_2010}
{Wanderman}, D., \& {Piran}, T. 2010, mnras, 406, 1944

\bibitem[{{Wang} {et~al.}(2015){Wang}, {Dai}, \& {Liang}}]{Wang}
{Wang}, F.~Y., {Dai}, Z.~G., \& {Liang}, E.~W. 2015, nar, 67, 1

\bibitem[{{Woosley} \& {Bloom}(2006)}]{Woosley2006}
{Woosley}, S.~E., \& {Bloom}, J.~S. 2006, araa, 44, 507

\bibitem[{{Xie} {et~al.}(2023){Xie}, {Nong}, {Zhang}, {Wang}, {Li}, \& {Liang}}]{Hanbei}
{Xie}, H., {Nong}, X., {Zhang}, B., {et~al.} 2023, arXiv e-prints, arXiv:2307.16467

\bibitem[{{Yonetoku} {et~al.}(2004){Yonetoku}, {Murakami}, {Nakamura}, {Yamazaki}, {Inoue}, \& {Ioka}}]{Yonetoku}
{Yonetoku}, D., {Murakami}, T., {Nakamura}, T., {et~al.} 2004, apj, 609, 935

\end{thebibliography}

\end{document}